\documentclass[pra,aps,superscriptaddress,footinbib,twocolumn]{revtex4-1}

\usepackage{mathrsfs}
\usepackage{amsfonts}
\usepackage{amssymb}
\usepackage{amsmath}
\usepackage{graphicx}
\usepackage[usenames,dvipsnames]{color}
\usepackage[colorlinks=true,citecolor=blue,linkcolor=magenta]{hyperref}
\usepackage{lmodern}
\usepackage{amsthm}
\usepackage{dsfont}
\usepackage{natbib}

\newcommand{\figpath}{.}

\newcommand{\Tr}{\mathrm{Tr}}

\newcommand{\absLR}[1]{\left\vert #1 \right\vert}

\newcommand{\ket}[1]{\vert{ #1 }\rangle}
\newcommand{\bra}[1]{\langle{ #1 }\vert}
\newcommand{\ketbra}[2]{\vert #1 \rangle \langle #2 \vert}

\newcommand{\mean}[1]{\langle #1 \rangle}

\newcommand{\bfzero}{\vec{\boldsymbol{0}}}

\begin{document}

\title{Dual-state purification for practical quantum error mitigation}

\author{Mingxia Huo}

\affiliation{Department of Physics and Beijing Key Laboratory for Magneto-Photoelectrical Composite and Interface Science, School of Mathematics and Physics, University of Science and Technology Beijing, Beijing 100083, China}

\author{Ying Li}
\email{yli@gscaep.ac.cn}
\affiliation{Graduate School of China Academy of Engineering Physics, Beijing 100193, China}

\begin{abstract}
Quantum error mitigation is essential for computing on the noisy quantum computer with a limited number of qubits. In this paper, we propose a practical protocol of error mitigation by virtually purifying the quantum state without qubit overhead or requiring only one ancillary qubit. In dual-state purification, we effectively generate a purified state with increased fidelity using the erroneous state and its dual state, respectively, prepared with the noisy quantum circuit and the dual map of its inverse circuit. Combined with tomography purification, we can make sure that the final estimate of an observable is obtained from a pure state. The numerical result suggests that our protocol reduces the error by a rescaling factor decreasing with the qubit number and circuit depth, i.e.~the performance of purification is better for larger circuits. On a cloud quantum computer, we successfully demonstrate the reduced error with a quantum variational eigensolver circuit. 
\end{abstract}

\maketitle

\section{Introduction}

The problem of errors caused by decoherence and imperfect control is one of the main obstacles to achieving practical quantum computing. With error rates lower than the fault-tolerance threshold and sufficient physical qubits, we can suppress errors on logical qubits to any level using quantum error correction~\cite{Knill1998, Raussendorf2007}. However, full-scale application of quantum error correction is unfeasible with noisy intermediate-scale quantum technologies due to the limited qubit number~\cite{Fowler2012, OGorman2017, Preskill2018}. Quantum error mitigation is an alternative way to suppress errors~\cite{Li2017, Temme2017, Endo2018, McClean2017}. Without the undesired qubit overhead, some quantum error mitigation protocols are immediately implementable on today's quantum computers~\cite{Kandala2019, Song2019, Zhang2020, Arute2020, Colless2018}. 

Several quantum error mitigation approaches have been developed. Most of them can be classified into three categories. The first category includes approaches based on knowledge of the error model, such as error extrapolation and probabilistic error cancellation~\cite{Li2017, Temme2017, Endo2018}. The second category includes approaches based on constraints on the error-free quantum state, such as symmetry verification~\cite{McArdle2019, Bonet2018} and purification of fermion correlations~\cite{Arute2020}. Other approaches developed for specific algorithms belong to the third category, for instance, subspace expansion~\cite{McClean2017}. Compared with the first category, the successful implementation of a constraint-based approach does not rely on error model benchmarking, e.g.~gate set tomography~\cite{Merkel2013, Stark2014, Greenbaum2015, BlumeKohout2017} and sampling trial circuits~\cite{Strikis2020, Czarnik2020, Wang2021}. Recently, virtual distillation protocols are proposed for error mitigation, exploring the universal constraint that the error-free state is a pure state~\cite{Koczor2020, Huggins2020, Czarnik2021}. In virtual distillation, multiple copies of the erroneous state are used for measuring an observable in the purified state. In this paper, we propose a purification protocol without the qubit overhead for storing copies of the state, and our protocol is practical for today's quantum computers. 

\begin{figure}[tbp]
\begin{center}
\includegraphics[width=1\linewidth]{\figpath/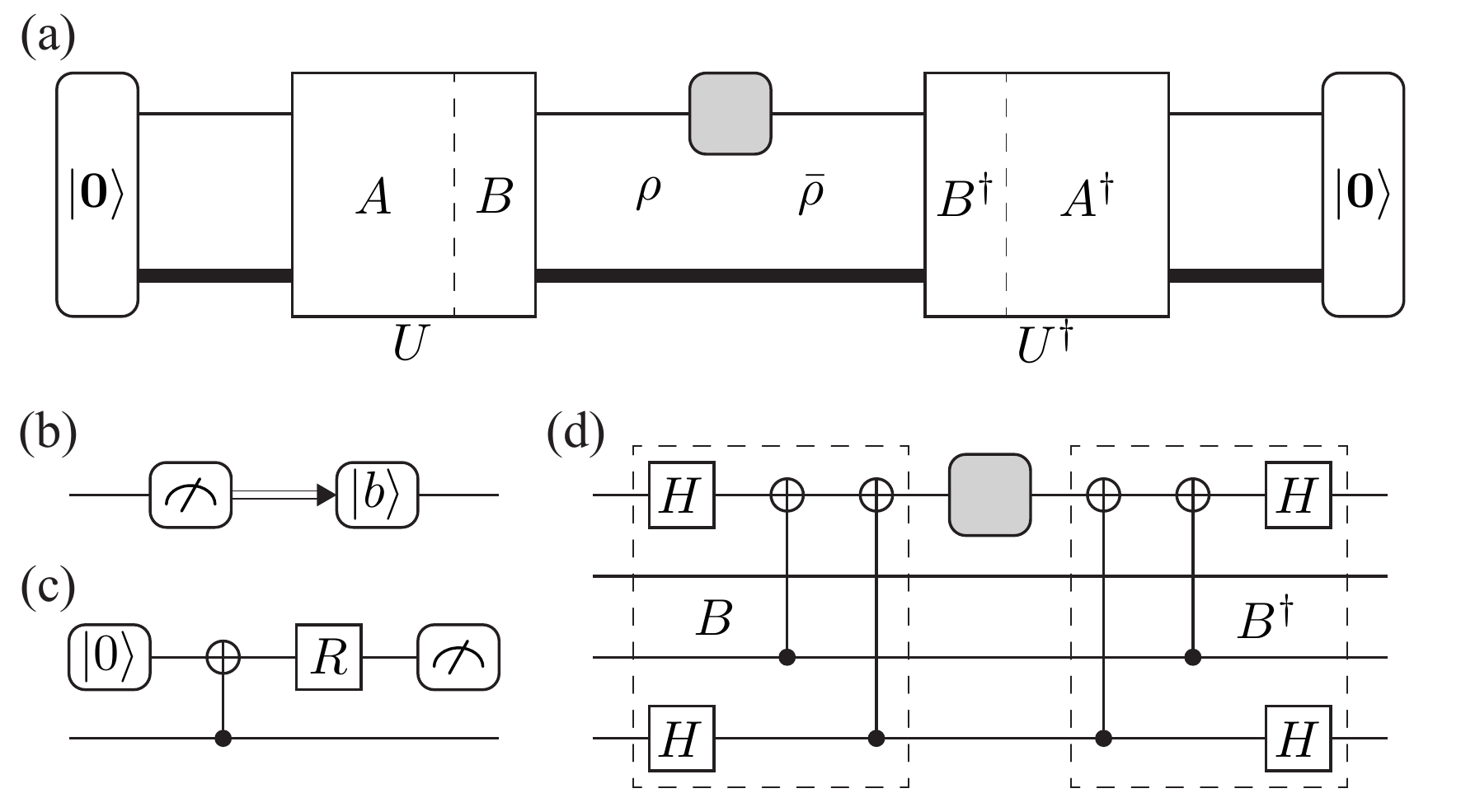}
\caption{
(a) Circuit of dual-state purification. (b) Intermediate measurement without ancillary qubit. (c) Intermediate measurement with one ancillary qubit. (d) Measurement basis transformation circuit. 
}
\label{fig:circuit}
\end{center}
\end{figure}

We propose two methods of purification. In dual-state purification, to promote the final state fidelity of a quantum circuit, we prepare a second copy of the state using the dual map of the inverse circuit. The entire circuit consists of the original circuit and its inverse circuit sandwiched with an intermediate measurement [see Fig.~\ref{fig:circuit}(a)]. The circuit-inverse-circuit structure has been proposed for error mitigation in verified phase estimation~\cite{OBrien2020}. When the direct intermediate measurement is not available on a quantum computer, such as \textit{ibmq} quantum computers, we can effectively implement the measurement with an ancillary qubit. If the entire circuit is error-free, at its end, the ancillary qubit is in a pure state. In tomography purification, we further suppress errors on top of dual-state purification by applying state tomography~\cite{Greenbaum2015} to the ancillary qubit and post-processing data according to the pure-state constraint. 

Our protocol is efficient in qubit overhead and error suppression. Because we prepare the state and dual state on the same qubits, we do not use extra qubits for storing copies of the state. Additionally, we do not use controlled-swap operations employed in virtual distillation protocols~\cite{Koczor2020, Huggins2020, Czarnik2021}. In our circuit for purification, all operations are part of state preparation except for the intermediate measurement. Dual-state purification suppresses errors in all these state-preparation operations. We note that the intermediate measurement is either a single-qubit measurement or accomplished by only one additional controlled-NOT gate. Tomography purification can eliminate remaining errors after dual-state purification, including those caused by the intermediate measurement. In the numerical simulation, we show that our protocol can reduce the error in the final observable by orders of magnitude, even when the intermediate measurement is noisy. The error rescaling factor decreases with the qubit number and circuit depth in the random circuit test, i.e.~the performance of purification is better for larger circuits. We experimentally implement our protocol on an \textit{ibmq} quantum computer, and we observe the error reduced by a rescaling factor of $0.105$. 

\section{Dual-state purification}

In purification using two copies of a state, the state is purified by taking $\rho \rightarrow \rho^2/\Tr\left(\rho^2\right)$. Given an observable $O$, we estimate its expected value with $\mean{O} = \Tr\left(O\rho^2\right)/\Tr\left(\rho^2\right)$. 

On a quantum computer, we prepare the state $\rho$ using a quantum circuit. Suppose the circuit realises a unitary transformation $U$, the final state without any error is $\ket{\psi} = U\ket{\bfzero}$, where $\ket{\bfzero}$ denotes that all qubits are initialised in the state $0$. When the final state is error-free, i.e.~$\rho = \ketbra{\psi}{\psi}$, we have the following equation: 
\begin{eqnarray}
\Tr\left(O\rho^2\right) &=& \Tr\left(U\ketbra{\bfzero}{\bfzero}U^\dag O U\ketbra{\bfzero}{\bfzero}U^\dag\right) \notag \\
&=& \bra{\bfzero}U^\dag O U\ketbra{\bfzero}{\bfzero}U^\dag U\ket{\bfzero}.
\label{eq:unitary}
\end{eqnarray}

On a noisy quantum computer, the implementation of a circuit is imperfect. We use the trace-preserving completely positive map $\mathcal{U}(\bullet) = \sum_k F_k \bullet F_k^\dag$ to represent the quantum process realised with the noisy circuit. Then, the final state with error is $\rho = \mathcal{U}(\ketbra{\bfzero}{\bfzero})$. When the fidelity of circuit is close to one, the noisy process is close to the unitary process, i.e.~$\mathcal{U}(\bullet) \approx U \bullet U^\dag$; therefore, $\rho \approx \ketbra{\psi}{\psi}$. 

Given a unitary circuit, we can always compose its inverse circuit, corresponding to the inverse transformation $U^\dag$. With the noise, the quantum process of the inverse circuit is $\mathcal{V}(\bullet) = \sum_k G_k \bullet G_k^\dag$. When the fidelity is close to one, we have $\mathcal{V}(\bullet) \approx U^\dag \bullet U$. 

Now, we replace $U \bullet U^\dag$ and $U^\dag \bullet U$ with $\mathcal{U}$ and $\mathcal{V}$, respectively, on the second line in Eq.~(\ref{eq:unitary}). We obtain the key formula of dual-state purification 
\begin{eqnarray}
&& \bra{\bfzero}\mathcal{V}\big(O\mathcal{U}\left(\ketbra{\bfzero}{\bfzero}\right)\big)\ket{\bfzero} \notag \\
&=& \Tr\left(\bar{\mathcal{V}}\left(\ketbra{\bfzero}{\bfzero}\right) O \mathcal{U}\left(\ketbra{\bfzero}{\bfzero}\right)\right) 
= \Tr\left(O \rho\bar{\rho}\right),
\end{eqnarray}
where $\bar{\mathcal{V}}(\bullet) = \sum_k G_k^\dag \bullet G_k$ is the dual map of $\mathcal{V}$, and we call the state $\bar{\rho} = \bar{\mathcal{V}}\left(\ketbra{\bfzero}{\bfzero}\right)$ the dual state of $\rho$~\cite{footnote1}. When the fidelity is close to one, $\bar{\mathcal{V}}(\bullet) \approx U \bullet U^\dag$ and $\bar{\rho} \approx \ketbra{\psi}{\psi}$. 

Although states $\rho$ and $\bar{\rho}$ are different, purification still works when $\ketbra{\psi}{\psi}$ is the dominant component in both $\rho$ and $\bar{\rho}$ with incoherent errors. We express erroneous states in the form $\rho = F\ketbra{\psi}{\psi} + p\rho_e$ and $\bar{\rho} = \bar{F}\ketbra{\psi}{\psi} + \bar{p}\bar{\rho}_e$, where fidelities $F$ and $\bar{F}$ are close to one, $p,\bar{p}\ll 1$, $\rho_e$ and $\bar{\rho}_e$ are normalised positive semidefinite states, and $\ketbra{\psi}{\psi}\rho_e = \ketbra{\psi}{\psi}\bar{\rho}_e = 0$ for incoherent errors. Then, the state after purification is 
\begin{eqnarray}
\frac{\rho\bar{\rho}+\bar{\rho}\rho}{2} = F\bar{F} \ketbra{\psi}{\psi} + p\bar{p}\frac{\rho_e\bar{\rho}_e+\bar{\rho}_e\rho_e}{2},~~
\end{eqnarray}
in which the error has been reduced from $p$ and $\bar{p}$ to $p\bar{p}$. We will show later that the fidelity close to one is not a necessary condition of dual-state purification. We take $(\rho\bar{\rho}+\bar{\rho}\rho)/2$ as the purified state such that its matrix is Hermitian. 

In dual-state purification, we estimate the expected value of an observable with 
\begin{eqnarray}
\mean{O} = \Tr\left(O\frac{\rho\bar{\rho}+\bar{\rho}\rho}{2}\right) \Big/ \Tr\left(\frac{\rho\bar{\rho}+\bar{\rho}\rho}{2}\right).
\label{eq:DSP}
\end{eqnarray}
Next, we show how to compute such a formula using quantum circuits. 

\section{Circuit without ancillary qubit}

The circuit of dual-state purification is shown in Fig.~\ref{fig:circuit}(a). Suppose the state $\rho$ is prepared on $n$ qubits, the upper thin line represents qubit-1 (it is not the ancillary qubit), and the lower thick line represents other $n-1$ qubits. First, these $n$ qubits are initialised in the state $\ket{\bfzero}$; Then, the circuit $U$ and inverse circuit $U^\dag$ sandwiched with an intermediate measurement (denoted by the gray box) are performed; Finally, qubits are measured in the computational basis, and only the outcome $\ket{\bfzero}$ is selected. We discuss the observable $O = Z_1$ (Pauli operator of qubit-1) first and the general case later. 

To compute the denominator in Eq.~(\ref{eq:DSP}), we set the intermediate measurement to be idle. The denominator is the probability of measurement outcome $\ket{\bfzero}$, i.e. 
\begin{eqnarray}
P_{\bfzero} = \bra{\bfzero}\mathcal{V}\big(\mathcal{U}\left(\ketbra{\bfzero}{\bfzero}\right)\big)\ket{\bfzero} = \Tr\left(\rho\bar{\rho}\right).
\end{eqnarray}

To compute the numerator in Eq.~(\ref{eq:DSP}), the intermediate measurement without ancillary qubit is shown in Fig.~\ref{fig:circuit}(b): Qubit-1 is measured in the computational basis, and then it is initialised according to the measurement outcome, i.e.~the measurement is (effectively) a projective measurement. The measurement operator of outcome $b = 0,1$ is $E_b = \left[\openone+(-1)^b Z_1\right]/2$. The joint probability of intermediate outcome $b$ and final outcome $\ket{\bfzero}$ is 
\begin{eqnarray}
\tilde{P}_{\bfzero,b} = \bra{\bfzero}\mathcal{V}\big(E_b\mathcal{U}\left(\ketbra{\bfzero}{\bfzero}\right)E_b\big)\ket{\bfzero} = \Tr\left(E_b \rho E_b \bar{\rho}\right).
\end{eqnarray}
We can find that the numerator is $\tilde{P}_{\bfzero,0} - \tilde{P}_{\bfzero,1}$. Therefore, the expected value of $Z_1$ is 
\begin{eqnarray}
\mean{O} = \left( \tilde{P}_{\bfzero,0} - \tilde{P}_{\bfzero,1} \right) / P_{\bfzero}.
\label{eq:OPro}
\end{eqnarray}

\begin{figure}[tbp]
\begin{center}
\includegraphics[width=1\linewidth]{\figpath/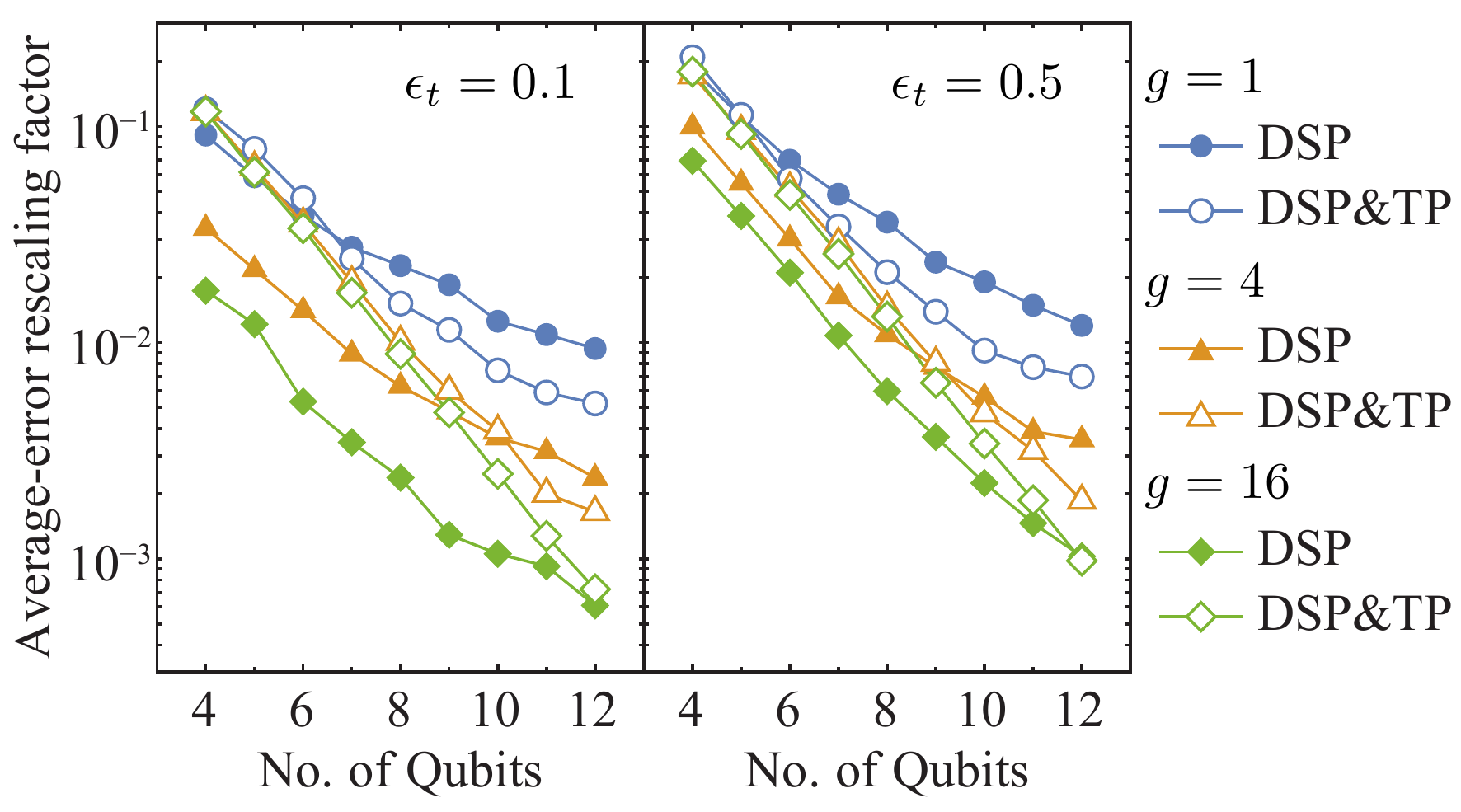}
\caption{
Average-error rescaling factors in the random circuit test. Dual-state purification (DSP) with the ancillary-qubit circuit and tomography purification (TP) are used to mitigate errors. We take gate numbers $n_G = gn^2$, and $n$ is the qubit number. Rescaling factors are not significantly changed when the total error rate increases from $\epsilon_t=0.1$ to $0.5$, i.e.~the fidelity close to one is not a necessary condition of DSP. 
}
\label{fig:rescaling}
\end{center}
\end{figure}

\section{Circuit with one ancillary qubit}

On some quantum computers, the measurement and reinitalisation are slow. In this case, we can replace the direct intermediate measurement with the circuit shown in Fig.~\ref{fig:circuit}(c): The upper line is the ancillary qubit, and the lower line is qubit-1. Taking $R$ as identity, the circuit effectively realises a projective measurement on qubit-1. The numerator is $\tilde{P}_{\bfzero,0} - \tilde{P}_{\bfzero,1} = \tilde{P}_{\bfzero} \mean{Z_a}_{\bfzero}$, where $\tilde{P}_{\bfzero} = \tilde{P}_{\bfzero,0} + \tilde{P}_{\bfzero,1}$ is the probability of $\ket{\bfzero}$ in the final measurement~\cite{footnote2}, and $\mean{Z_a}_{\bfzero} = (\tilde{P}_{\bfzero,0} - \tilde{P}_{\bfzero,1})/\tilde{P}_{\bfzero}$ is the expected value of ancillary-qubit Pauli operator $Z_a$ conditioned on the final outcome $\ket{\bfzero}$. 

The denominator can also be measured using the circuit in Fig.~\ref{fig:circuit}(c). Taking $R=H$, we effectively change the measurement basis of ancillary qubit from $Z$ to $X$. When the measurement outcome is $\ket{+}_a$, the controlled-NOT gate becomes trivial and does not have any effect on qubit-1. Therefore, the denominator is the probability of $\ket{\bfzero}$ conditioned on the intermediate outcome $\ket{+}_a$, i.e.~$P_{\bfzero} = \tilde{P}_{\bfzero}\left(1+\mean{X_a}_{\bfzero}\right)$~\cite{footnote3}. Here, $\mean{X_a}_{\bfzero}$ is the expected value of ancillary-qubit Pauli operator $X_a$ conditioned on the final outcome $\ket{\bfzero}$. We remark that changing $R$ does not change $\tilde{P}_{\bfzero}$, i.e.~the probability of $\ket{\bfzero}$ in the final measurement. 

With the ancillary-qubit circuit, the expected value of $Z_1$ is 
\begin{eqnarray}
\mean{O} = \mean{Z_a}_{\bfzero} / \left(1+\mean{X_a}_{\bfzero}\right).
\label{eq:OPauli}
\end{eqnarray}
To compute $\mean{O}$, we only need to measure expected values of ancillary-qubit Pauli operators conditioned on the final outcome $\ket{\bfzero}$. Using post-selected data to estimate the observable makes tomography purification possible. Later, we will show that tomography purification is critical in the experimental implementation on the \textit{ibmq} quantum computer. 

Only errors in the intermediate measurement [shown in Figs.~\ref{fig:circuit}(b) and (c)] cannot be suppressed in dual-state purification. Let $\rho_{\bfzero}$ and $E_{\bfzero}$ be the $n$-qubit initial state and final measurement operator with state preparation and measurement errors, respectively. Then, the state $\rho$ and its dual state $\bar{\rho}$ become $\rho = \mathcal{U}\left(\rho_{\bfzero}\right)$ and $\bar{\rho} = \bar{\mathcal{V}}\left(E_{\bfzero}\right)$, respectively. Note that gate errors have been taken into account in $\mathcal{U}$ and $\bar{\mathcal{V}}$. All these errors are suppressed in dual-state purification as long as they are incoherent. We can suppress errors in the intermediate measurement by tomography purification as we show next. 

\section{Tomography purification}

There are two constraints on the state of ancillary qubit. First, when the entire circuit is error-free, the final state of $n+1$ qubits is a pure state. Therefore, when the $n$ qubits are projected onto $\ket{\bfzero}$ by the final measurement, the state of ancillary qubit is a pure state. Second, the expected value of Pauli operator $Y_a$ is $\mean{Y_a}_{\bfzero} = 0$ if the entire circuit is error-free (see Appendix~\ref{app:EFS}). Here, we only explore the first constraint. 

We can eliminate errors on the ancillary qubit according to the pure-state constraint using tomography purification. First, instead of only measuring in the $X$ and $Z$ bases, we perform a full tomography~\cite{Greenbaum2015} on the ancillary qubit to obtain its state $\rho_{a\vert\bfzero}$ conditioned on the final outcome $\ket{\bfzero}$. Then, we calculate the eigenstate of $\rho_{a\vert\bfzero}$ with the larger eigenvalue, denoted by $\ket{\chi}_a$. Finally, we compute expected values of $X$ and $Z$ in the state $\ket{\chi}_a$, i.e.~$\mean{X_a}_{\bfzero} = \bra{\chi}_a X_a \ket{\chi}_a$ and $\mean{Z_a}_{\bfzero} = \bra{\chi}_a Z_a \ket{\chi}_a$. In this way, we purify the state of ancillary qubit. We remark that the ancillary qubit is in a mixed state due to not only errors directly occurring on it but also errors on other $n$ qubits teleported to the ancillary qubit by the projection onto $\ket{\bfzero}$. 

\section{General observable}

The observable $O$ can always be expressed as a linear combination of $n$-qubit Pauli operators $\sigma$, i.e~$O = \sum_{\sigma} \alpha_\sigma \sigma$. Accordingly, we can obtain the expected value of $O$ by measuring each Pauli operator in the summation. For each $\sigma$ (except identity), there is a unitary operator $B$ transforming $\sigma$ into $Z_1$, i.e.~$B \sigma B^\dag = Z_1$. We can indirectly measure $\sigma$ using the circuit in Fig.~\ref{fig:circuit}(a): When we want to compute $\bra{\phi}\sigma\ket{\phi}$, where $\ket{\phi} = A\ket{\bfzero}$, we take $U = BA$, then we have $\bra{\phi}\sigma\ket{\phi} = \bra{\psi}Z_1\ket{\psi}$. Here, $A$ can be any unitary operator realised using quantum gates. In Appendix~\ref{app:MBT}, we give two protocols of implementing $B$ using Clifford gates on all-to-all and linear qubit networks, respectively. In Fig.~\ref{fig:circuit}(d), we give a circuit of the operator $B$ for $\sigma = X_1Z_3X_4$ as an example (qubits are sorted from top to bottom). 

\begin{figure*}[tbp]
\begin{center}
\includegraphics[width=1\linewidth]{\figpath/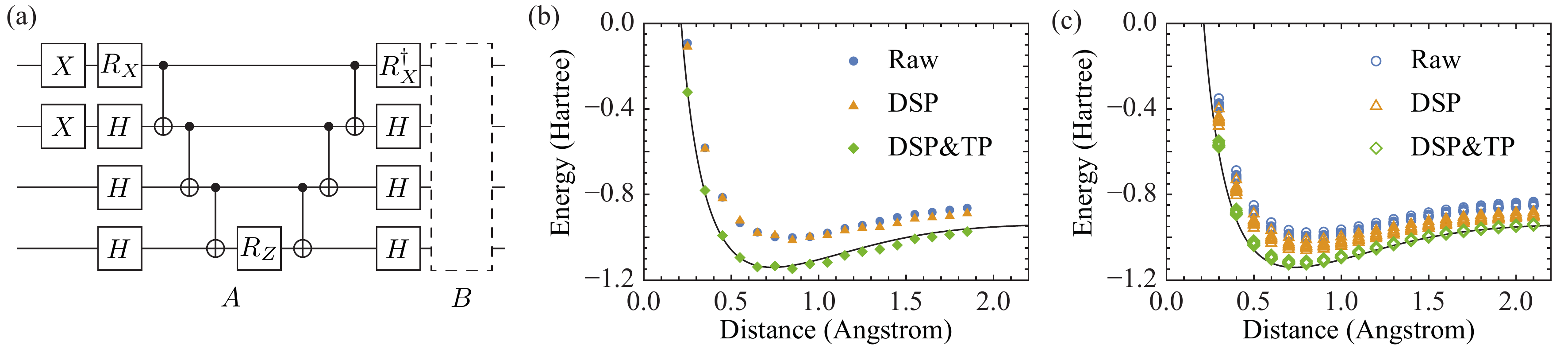}
\caption{
(a) Variational circuit. Gates $R_X = e^{i\frac{\pi}{4}Z}$ and $R_Z = e^{i\frac{\theta}{2}Z}$. (b) Ground state energy computed on the cloud quantum computer \textit{ibmq{\_}athens}. Taking Hamiltonian as the observable in Eq.~(\ref{eq:AER}), the average-error rescaling factor is $0.105$. Raw values are computed without error mitigation, and their average error is $0.133$. Dual-state purification (DSP) with the ancillary-qubit circuit and tomography purification (TP) are used to mitigate errors. DSP reduces the average error to $0.121$, and TP further reduces the average error to $0.014$. (c) Ground state energy computed on the numerically simulated quantum computer. Ten different error models are randomly generated and used to produce data in the plot; see Appendix~\ref{app:CEM} for details of the error model generation. The average error rate per two-qubit gate and measurement is $0.02$. 
}
\label{fig:VQE}
\end{center}
\end{figure*}

\section{Scaling behaviour and random circuit test}

Now, we discuss the performance of state purification when the qubit number and circuit depth increase. We consider the equal-error-probability state in the form $\rho = \bar{\rho} = F\ketbra{\psi}{\psi} + \frac{p}{M} \sum_{m=1}^M \ketbra{\psi^\perp_m}{\psi^\perp_m}$, where $\ket{\psi}$ and $\ket{\psi^\perp_m}$ are orthonormal eigenstates of $\rho$. For such a state, the fidelity of purified state $\rho^2/\Tr\left(\rho^2\right)$ is $F^2/\left(F^2 + p^2/M\right) \simeq 1 - p^2/(MF^2)$. Purification reduces the error probability from $p$ to $\sim p^2/(MF^2)$. We can find that the error rescaling factor $\sim p/(MF^2)$ decreases with $M$: The performance of purification is better when the total error probability $p$ is distributed among more eigenstates. Eigenstates $\ket{\psi^\perp_m}$ are created by errors in quantum gates, and more gates create more eigenstates. Therefore, we expect that the error rescaling factor decreases with the gate number. The other factor is the Hilbert space dimension. When the dimension is higher, it is more unlikely that two errors create the same eigenstate (i.e.~one of them does not increase $M$). Therefore, we expect that the error rescaling factor also decreases with the qubit number. 

In the numerical simulation of randomly generated circuits, we show that the error rescaling factor decreases with the qubit number and circuit depth, which coincides with conclusions derived from the equal-error-probability state. We randomly generate circuits with $n=4,5,\ldots,12$ qubits. Circuits are formed of controlled-NOT gates and single-qubit gates. For each $n$, we take $n_G = n^2,4n^2,16n^2$ as the number of controlled-NOT gates for the transformation $A$ [see Fig.~\ref{eq:OPro}(a)]. The observable $O$ is a randomly selected $n$-qubit Pauli operator, which determines the transformation $B$. The depolarising error model~\cite{Knill2002} is used in the numerical simulation. Error rates of controlled-NOT gates and measurements are around the average error rate $\epsilon = \epsilon_t/n_G$ with $50\%$ fluctuation. Here $\epsilon_t \sim 1-F$ is the total gate error rate. Errors in single-qubit gates are neglected. See Appendix~\ref{app:RCT} for further details of random circuits. 

We evaluate the average-error rescaling factor in the random circuit test. For each $(n,n_G,\epsilon_t)$, we generate $100$ random circuits. For each circuit, we compute the error-free expected value $\mean{O}_{ef}$, the value produced by noisy circuit without error mitigation $\mean{O}_n$, and the value after error mitigation $\mean{O}_{em}$. Then, we calculate the average-error rescaling factor 
\begin{eqnarray}
r = \frac{\mean{\absLR{\mean{O}_{em} - \mean{O}_{ef}}}_{c}}{\mean{\absLR{\mean{O}_{n} - \mean{O}_{ef}}}_{c}},
\label{eq:AER}
\end{eqnarray}
where $\mean{\bullet}_{c}$ denotes the average over circuits for test, i.e.~$100$ random circuits in this case. The rescaling factor is plotted in Fig.~\ref{fig:rescaling}. 

In Fig.~\ref{fig:rescaling}, we can find that the average-error rescaling factor decreases with the qubit number and circuit depth for both total error rates $\epsilon_t = 0.1,0.5$. The error mitigation with tomography purification is superior only when the qubit number is large, i.e.~there is a cross between two curves with markers in the same shape. When there are more qubits in the circuit, errors teleported to the ancillary qubit may be closer to depolarising errors. Tomography purification is ideal for dealing with depolarising errors on the ancillary qubit, which may explain its advantage for large qubit numbers. 

\section{Demonstration on quantum computer}

On the cloud quantum computer \textit{ibmq{\_}athens}, we apply dual-state purification to the circuit in Fig.~\ref{fig:VQE}(a). According to the variational quantum eigensolver algorithm~\cite{Peruzzo2014}, this circuit is used to compute the ground state energy of ${\rm H}_2$ in the STO- 3G basis (see Appendix~\ref{app:VQE} and Ref.~\cite{McArdle2020}). In our demonstration, given a nuclear separation, we find the optimal value of the variational parameter $\theta$ on a classical computer, and then we use the optimal value to compute the ground state energy on the quantum computer. The result is plotted in Fig.~\ref{fig:VQE}(b). 

On \textit{ibmq{\_}athens}, dual-state purification incorporating with tomography purification can significantly improve the accuracy of quantum computing. However, dual-state purification itself does not work well. This experimental result is consistent with the numerical result of the composite error model that includes not only Pauli errors but also amplitude damping errors [see Fig.~\ref{fig:VQE}(c)]. The amplitude damping process can create coherent errors, which are not corrected by dual-state purification and teleported to the ancillary qubit. In this case, tomography purification becomes necessary. 

\section{Conclusions}

In this paper, we propose a practical purification protocol for quantum error mitigation. We successfully suppress the computational error in the variational quantum eigensolver circuit on a cloud quantum computer using our protocol. Although the error suppression observed in the experiment is due to tomography purification, dual-state purification is necessary. We can apply the state tomography directly to all of the $n$ qubits, which is, however, not scalable; with dual-state purification, we only apply the tomography to the ancillary qubit. 

Dual-state purification is better at correcting Pauli errors than general errors such as amplitude damping, as shown in the random circuit test and experimental result. Therefore, Pauli twirling, which converts general errors into Pauli errors, may be necessary for certain machines (i.e.~error models). The error rescaling factor decreases rapidly with the qubit number in the random circuit test. If this trend persists to tens of qubits, dual-state purification will be a simple and efficient way to attain high-accuracy quantum computing with noisy intermediate-scale quantum technologies. 

\begin{acknowledgments}
We acknowledge the use of simulation toolkit QuESTlink~\cite{Jones2020} and IBM Quantum services~\cite{IBMQuantum} for this work. 
This work is supported by National Natural Science Foundation of China (Grant No. 11574028, 11874083 and 11875050). YL is also supported by NSAF (Grant No. U1930403). 
\end{acknowledgments}

\appendix

\section{Error-free states}
\label{app:EFS}

We suppose the entire circuit is error-free. After the unitary transformation $U$, the state of $n+1$ qubits is $\ket{\psi}\otimes\ket{0}_a$. After the controlled-NOT gate, the state becomes 
\begin{eqnarray}
\frac{1}{\sqrt{2}} \left( \ket{\psi}\otimes\ket{+}_a + Z_1\ket{\psi}\otimes\ket{-}_a \right).
\end{eqnarray}
The inverse transformation $U^\dag$ and the final projective measurement onto $\ket{\bfzero}$ are equivalent to a projective measurement onto $\ket{\psi}$, i.e.~the final state of the ancillary qubit is 
\begin{eqnarray}
\ket{\chi}_a = \frac{1}{\sqrt{2\tilde{P}_{\bfzero}}} \left( \ket{+}_a + \bra{\psi}Z_1\ket{\psi}\ket{-}_a \right),
\end{eqnarray}
where 
\begin{eqnarray}
\tilde{P}_{\bfzero} = \frac{1+\bra{\psi}Z_1\ket{\psi}^2}{2} \geq \frac{1}{2}.
\end{eqnarray}

The expected values of Pauli operators in the final state of ancillary qubit are 
\begin{eqnarray}
\mean{X_a}_{\bfzero} &=& \bra{\chi}_a X_a \ket{\chi}_a = \frac{1-\bra{\psi}Z_1\ket{\psi}^2}{2\tilde{P}_{\bfzero}} = \frac{1}{\tilde{P}_{\bfzero}} - 1, \notag \\
\mean{Y_a}_{\bfzero} &=& \bra{\chi}_a Y_a \ket{\chi}_a = 0, \notag \\
\mean{Z_a}_{\bfzero} &=& \bra{\chi}_a Z_a \ket{\chi}_a = \frac{\bra{\psi}Z_1\ket{\psi}}{\tilde{P}_{\bfzero}}.
\end{eqnarray}
Here, we have used that $\bra{\psi}Z_1\ket{\psi}$ is real. 

\section{Measurement basis transformation}
\label{app:MBT}

We can express an $n$-qubit Pauli operator as 
\begin{eqnarray}
\sigma = \prod_{i=1}^n P_i,
\end{eqnarray}
where $P = I,X,Y,Z$, and $P_i$ is the Pauli operator on qubit-$i$. To realise the measurement basis transformation $B$, we first apply single-qubit gates $R_i$ on each qubit, 
\begin{eqnarray}
R_i = \left\{
\begin{array}{ll}
I_i, & P_i = I_i,Z_i; \\
H_i, & P_i = X_i; \\
H_iS_i^3, & P_i = Y_i.
\end{array}
\right.
\end{eqnarray}
Here, $H$ is the Hadamard gate, and $S$ is the $\frac{\pi}{2}$ phase gate. These single-qubit gates transform $\sigma$ into 
\begin{eqnarray}
\sigma' = B_1 \sigma B_1^\dag = \prod_{i=1}^n P_i',
\end{eqnarray}
where 
\begin{eqnarray}
B_1 = \prod_{i=1}^n R_i
\end{eqnarray}
and 
\begin{eqnarray}
P_i' = \left\{
\begin{array}{ll}
I_i, & P_i = I_i; \\
Z_i, & P_i = X_i,Y_i,Z_i.
\end{array}
\right.
\end{eqnarray}
Single-qubit Pauli operators in $\sigma'$ are either $I$ or $Z$. 

On an all-to-all qubit network, we can apply the controlled-NOT gate $\Lambda_{X,i,j}$ on any pair of qubits $i$ and $j$. Here, $i$ is the control qubit, and $j$ is the target qubit. Without loss of generality, we suppose $P_1' = Z_1$; If it is not the case, just find any qubit with $P_i'=Z_i$ and let this qubit play the role of qubit-1. We transform $\sigma'$ into $Z_1$ by applying controlled-NOT gates 
\begin{eqnarray}
B_2 = \Lambda_n \cdots \Lambda_3 \Lambda_2,
\end{eqnarray}
where 
\begin{eqnarray}
\Lambda_i = \left\{
\begin{array}{ll}
\openone, & P_i' = I_i; \\
\Lambda_{X,i,1}, & P_i' = Z_i.
\end{array}
\right.
\end{eqnarray}
On an all-to-all qubit network, the overall transformation is $B = B_2 B_1$. 

On a linear qubit network, we can only apply the controlled-NOT gate on a pair of nearest-neighboring qubits $i$ and $i+1$. Then, the transformation from $\sigma'$ to $Z_1$ is realised by 
\begin{eqnarray}
B_2' = \Lambda_1' \Lambda_2' \cdots \Lambda_{m-1}',
\end{eqnarray}
where $P_m' = Z_m$, $P_i' = I_i$ for all $i>m$, and 
\begin{eqnarray}
\Lambda_i' = \left\{
\begin{array}{ll}
\Lambda_{X,i+1,i}\Lambda_{X,i,i+1}, & P_i' = I_i; \\
\Lambda_{X,i+1,i}, & P_i' = Z_i.
\end{array}
\right.
\end{eqnarray}
On a linear qubit network, the overall transformation is $B = B_2' B_1$. 

\section{Random circuit test}
\label{app:RCT}

Random circuits are generated on an all-to-all network. Given $n$ and $n_G$, we generate the circuit of transformation $A$ as follows: i) Right after the initialisation, apply a single-qubit gate on each qubit; ii) Randomly select two qubits and apply the controlled-NOT gate, then apply two single-qubit gates on the same qubits; iii) Repeat step-ii until $n_G$ controlled-NOT gates are applied. Single-qubit gates are uniformly sampled from the unitary group according to the Haar measure. 

The observable is $O=\prod_{i=1}^n P_i$, where $P_i=I_i,Z_i$ is the Pauli operator on the qubit-$i$. Because of random single-qubit gates in the circuit of $A$ (the last layer effectively changes the measurement basis), taking $P_i=Z_i$ is equivalent to taking $P_i=X_i,Y_i,Z_i$ with the same probability. Because controlled-NOT gates are randomly generated on the all-to-all network, we always take $P_1 = Z_1$ without loss of generality. Each $P_i$ with $i=2,3,\dots,n$ is drawn from $\{I_i,Z_i\}$ with the uniform distribution. 

The map of two-qubit depolarising error reads 
\begin{eqnarray}
\mathcal{E}_{i,j}\left(\epsilon_{i,j}\right) = (1-\frac{16}{15}\epsilon_{i,j})[\openone] + \frac{\epsilon_{i,j}}{15}\sum_{P_i,P_j} [P_i P_j],
\end{eqnarray}
where $P_i=I_i,X_i,Y_i,Z_i$, $P_j=I_j,X_j,Y_j,Z_j$, $\epsilon_{i,j}$ is the error rate of the two-qubit depolarising error on qubits $i$ and $j$, and $[U](\bullet) = U\bullet U^\dag$. Then, the controlled-NOT gate with depolarising error is $\mathcal{E}_{i,j}\left(\epsilon_{i,j}\right)[\Lambda_{X,i,j}]$. The error rate of measurement on qubit-$i$ is $\epsilon_{i,i}$: The measurement reports an incorrect outcome with the probability $\epsilon_{i,i}$. The average error rate is $\epsilon = \epsilon_t/n_G$. We randomly generate $\epsilon_{i,j}$ and $\epsilon_{i,i}$ from the uniform distribution in $[0.5\epsilon, 1.5\epsilon]$. 

To test the state purification protocols, for each $(n,n_G,\epsilon_t)$, we generate $100$ random circuits. Each random circuit is specified by a circuit of $A$, an observable $O$ and an error rate matrix $\epsilon_{i,j}$. Given the random circuit, we compute $\mean{O}_{ef}$, $\mean{O}_{n}$ and $\mean{O}_{em}$ as follows. To compute $\mean{O}_{ef}$, we simulate the circuit of $A$ with the error channels switched off and calculate the expected value. To compute $\mean{O}_{n}$, we simulate the circuit of $A$ with the error channels switched on and calculate the expected value. There are two protocols of error mitigation. In the dual-state purification protocol, first we work out the measurement basis transformation $B$ of the observable $O$ according to the protocol for all-to-all network (see Appendix~\ref{app:MBT}), then we simulate the ancillary-qubit circuit of dual-state purification with the error channels switched on and compute the expected value $\mean{O}_{em}$. In the dual-state plus tomography purification protocol, the circuit is the same, the ancillary-qubit is measured in $X$, $Y$ and $Z$ bases for tomography, and then we compute the expected value $\mean{O}_{em}$ according to tomography purification. 

\section{Variational quantum eigensolver}
\label{app:VQE}

The STO-3G basis includes 4 spin-orbitals, which can be encoded into four qubits. The corresponding Hamiltonian of qubits reads 
\begin{eqnarray}
H &=& h_{0}I+h_{1}Z_{0}+h_{2}Z_{1}+h_{3}Z_{2}+h_{4}Z_{3} \notag \\
&& +h_{5}Z_{1}Z_{0}+h_{6}Z_{2}Z_{0}+h_{7}Z_{3}Z_{0} \notag \\
&& +h_{8}Z_{2}Z_{1}+h_{9}Z_{3}Z_{1}+h_{10}Z_{3}Z_{2} \notag \\
&& + h_{11}X_{3}X_{2}Y_{1}Y_{0}+h_{12}Y_{3}Y_{2}X_{1}X_{0} \notag \\
&& +h_{13}X_{3}Y_{2}Y_{1}X_{0}+h_{14}Y_{3}X_{2}X_{1}Y_{0}.
\end{eqnarray}
We use \textit{Qiskit} to calculate the coefficients $h_k$ in the Hamiltonian. The variational circuit is the simplified UCCSD-circuit given in Ref.~\cite{McArdle2020}, which has only one variational parameter $\theta$. Suppose $\ket{\phi(\theta)}$ is the final state of the variational circuit, we find the optimal value of $\theta$ by minimising the expected value of energy $\bra{\phi(\theta)} H \ket{\phi(\theta)}$. 

The variational circuit is the circuit $A$. For each Pauli operator in $H$, we compose the corresponding measurement basis transformation circuit $B$ according to the protocol for linear network (see Appendix~\ref{app:MBT}). 

\section{Composite error model}
\label{app:CEM}

In the error model including Pauli errors and amplitude damping, the error rate $\epsilon = 0.02$ is equally distributed to Pauli errors and amplitude damping errors. The map of a controlled-NOT gate with error reads 
\begin{eqnarray}
\mathcal{A}_{j}(\delta')\mathcal{A}_{i}(\delta)\mathcal{D}_{j}(\eta')\mathcal{D}_{i}(\eta)\mathcal{E}_{i,j}(\epsilon')[\Lambda_{X,i,j}], \notag
\end{eqnarray}
where 
\begin{eqnarray}
\mathcal{D}_{i}(\eta) = (1-\eta)[\openone] + \eta [Z_i]
\end{eqnarray}
is the dephasing error on qubit-$i$, and 
\begin{eqnarray}
\mathcal{A}_{i}(\delta) &=& \left[\frac{\openone+Z_i}{2} + \sqrt{1-\delta}\frac{\openone-Z_i}{2}\right] \notag \\
&&+ \left[\sqrt{\delta}\frac{X_i+iY_i}{2}\right]
\end{eqnarray}
is the amplitude damping on qubit-$i$. For each pair of qubits $(i,j)$, we randomly choose error rates according to the uniform distribution in the intervals $\epsilon',\eta,\eta'\in [0.5\epsilon/6,1.5\epsilon/6]$ and $\delta,\delta'\in [0.5\epsilon/2,1.5\epsilon/2]$. The measurement with error is modeled as applying the map 
$$\mathcal{A}_{i}(\delta)\mathcal{E}_{i}(\epsilon')$$
before the error-free measurement, where 
\begin{eqnarray}
\mathcal{E}_{i}\left(\epsilon'\right) = (1-\epsilon')[\openone] + \epsilon'\sum_{P_i=X_i,Y_i,Z_i} [P_i]
\end{eqnarray}
is the depolarising error on qubit-$i$. For each qubit $i$, we randomly choose error rates according to the uniform distribution in the intervals $\epsilon'\in [0.5\times 3\epsilon/4,1.5\times 3\epsilon/4]$ and $\delta\in [0.5\epsilon,1.5\epsilon]$.


\begin{thebibliography}{9}
\bibitem{footnote1} The dual map $\bar{\mathcal{V}}$ is always completely-positive, however, it is trace-preserving if and only if $\mathcal{V}$ is unital. Therefore, $\bar{\rho}$ is always positive semidefinite but may not be normalised.
\bibitem{footnote2} $\tilde{P}_{\bfzero}$ is different from $P_{\bfzero}$ because of the controlled-NOT gate on ancillary qubit.
\bibitem{footnote3} The joint probability of the intermediate outcome $\ket{+}_a$ and final outcome $\ket{\bfzero}$ is $\tilde{P}_{\bfzero}\left(1+\mean{X_a}_{\bfzero}\right)/2$, and the probability of $\ket{+}_a$ regardless of the final outcome is $1/2$.
\end{thebibliography}
\end{document}